\documentclass[fleqn,usenatbib,useAMS,letters]{mnras}

\usepackage{graphicx}	
\usepackage{amsmath}	
\usepackage{amssymb}	
\usepackage{multicol}        
\usepackage{bm}		
\usepackage{pdflscape}	
\usepackage{svg}
\usepackage{tikz}

\usepackage[T1]{fontenc}
\usepackage{ae,aecompl}

\usepackage{newtxtext,newtxmath}

\title[Follow the Water]{Follow the Water: Finding Water, Snow and Clouds on Terrestrial Exoplanets with Photometry and Machine Learning}

\author[Pham \& Kaltenegger]{
Dang Pham$^{1}$\thanks{E-mail: dang.pham@astro.utoronto.ca (DP)},
Lisa Kaltenegger$^{2}$
\\
$^{1}$David A. Dunlap Department of Astronomy \& Astrophysics, University of Toronto, 50 St. George Street, Toronto, ON M5S 3H4, Canada \\
$^{2}$Department of Astronomy and Carl Sagan Institute, Cornell University, 302 Space Sciences Building, Ithaca, NY 14853, USA
}

\date{Last updated YYYY MM DD; in original form YYYY MM DD}

\pubyear{2022}

\begin{document}
\label{firstpage}
\pagerange{\pageref{firstpage}--\pageref{lastpage}}
\maketitle

\begin{abstract}
All life on Earth needs water. NASA's quest to follow the water links water to the search for life in the cosmos. Telescopes like JWST and mission concepts like HabEx, LUVOIR and Origins are designed to characterise rocky exoplanets spectroscopically. However, spectroscopy remains time-intensive and therefore, initial characterisation is critical to prioritisation of targets.

Here, we study machine learning as a tool to assess water's existence through broadband-filter reflected photometric flux on Earth-like exoplanets in three forms: seawater, water-clouds and snow; based on 53,130 spectra of cold, Earth-like planets with 6 major surfaces. \texttt{XGBoost}, a well-known machine learning algorithm, achieves over 90\% balanced accuracy in detecting the existence of snow or clouds for S/N$\gtrsim 20$, and 70\% for liquid seawater for S/N $\gtrsim 30$. Finally, we perform mock Bayesian analysis with Markov-chain Monte Carlo with five  filters identified to derive exact surface compositions to test for retrieval feasibility.

The results show that the use of machine learning to identify water on the surface of exoplanets from broadband-filter photometry provides a promising initial characterisation tool of water in different forms. Planned small and large telescope missions could use this to aid their prioritisation of targets for time-intense follow-up observations.

\end{abstract}


\section{Introduction} \label{sec:intro}

Thousands of exoplanets have been detected to date; including terrestrial planets. Dozens of these orbit in their host star's habitable zone (HZ) \citep[e.g.][]{Kane2016, Johns2018}, where liquid water could be available on their surface \citep{Kasting1993}. Missions, such as the Transiting Exoplanet Survey Satellite \citep{Ricker2014}, and ground-based searches continually expand our list of potentially habitable planets \citep[e.g.][]{Nutzman2008, Luque2019, KalteneggerTESS2021}. Upcoming telescopes, such as the James Webb Space Telescope, will probe the atmosphere of Earth-sized exoplanets in the HZ \citep[e.g.][]{Zhang2018, KrissansenTotton2018, deWit2018, Wunderlich2019,Kaltenegger2020b,Edwards2021}.

Planetary formation models suggest that water-rich planets are common \citep{Raymond2013}, making planets with surface water interesting targets for upcoming telescope observations \citep[e.g.][]{Wunderlich2019, Smith2020a}. However, being in the HZ does not imply habitability \citep{Selsis2007, Kaltenegger2017}. Understanding the habitable characteristics of a planet requires a detailed characterisation of the exoplanet through time intense spectroscopic observations to analyze the planets' atmosphere and surface composition. But spectroscopy of Earth-size planets in the HZ will be time-intensive, even with future telescope concepts recommended by the recent Decadal survey like HabEx or LUVOIR \citep[e.g.][]{Feng2018, LUVOIR2019, Gaudi2020}. Thus, fast identification and prioritisation of targets is a critical aspect to enable a successful search. Several teams suggested photometry as a tool for such initial characterisation, testing it on Solar System objects as well as exoplanet models and explored optimal filters for such characterisation \citep[e.g.][]{Crow2011, Hegde2015, KrissansenTotton2016, Madden2018, Smith2020b}.

Additionally, machine learning methods have shown potential for fast initial characterisation of extrasolar planets with photometry: For giant planets \citet{Batalha2018} explore how to characterise atmospheric properties, such as metallicity and cloud coverage through supervised machine learning for the 5 broadband filters of the Nancy Grace Roman Space Telescope. For Earth-like planets, \citet{PhamKaltenegger2021} assess eight machine learning methods to characterise surface biota through photometric fluxes for the standard Johnson filters. 

Broadband filter photometry could allow smaller telescopes like the Nancy Grace Roman Telescope \citep{Spergel2015}, Ariel \citep{Tinetti2021} and Twinkle \citep{Edwards2019} to help initially characterising nearby rocky exoplanets. It could also allow large missions concepts like HabEx \citep{Gaudi2020} and LUVOIR \citep{LUVOIR2019} rapid initial characterisation of exoplanets in the Habitable Zone to prioritise targets for time-intense spectroscopic observations. This approach is complimentary to transit observations proposed for rocky exoplanet from JWST \citep{Gardner2006}, which won't be able to probe the surface of a rocky exoplanet.

In this article, we explore the capability of machine learning to detect water on the surface of a cool terrestrial planet in three forms: liquid seawater, water-clouds and snow. Note that we did not choose specific filters for any proposed telescope concept here, but we explore what the optimal choice would be to find water on an Earth-like exoplanet's surface to inform open design choices. 

We propose machine learning as an effective method to characterise terrestrial exoplanets' reflected broadband photometry, along with other statistical methods (e.g. compared to MCMC as shown in this paper). Our machine learning method \texttt{XGBoost} performs on a model grid of 53,130 Earth-size planets with varying surface coverage of six major surfaces. Our aims are to i) assess machine learning's efficacy in classifying the existence of water, clouds or snow, ii) identify optimal filters for characterisation for telescope designs and iii) perform mock Bayesian analysis on fluxes attained from these optimal filters.

First, we define an initial set of idealised, broadband filters. We then train a well-known, versatile machine learning model -- \texttt{XGBoost} \citep{Chen2016} -- with fluxes derived using the broadband filters, and assess it's performance at various S/N. This approach delivers an estimated signal-to-noise ratio cutoff for performance maximisation. In addition, the trained \texttt{XGBoost} identifies a set of five\footnote{We chose five filters based on previous instrument designs like the Nancy Grace Roman Space Telescope \citep{Spergel2013}. The methodology here can be easily extended to assess the trade-offs for  more (or less) filters.} optimal filters through feature importance ranking. Using the five optimal filters, we perform Bayesian inference through Markov-chain Monte Carlo sampling. The analysis described in this paper can be extended to any detected exoplanet by training the algorithm on specific models pertaining to the specific planetary environment (see \citealt{Truitt2020}).

This article is structured as follows: in section \ref{sec:data_gen}, we introduce the planetary models; in section \ref{sec:xgboost}, we discuss the usage of \texttt{XGBoost} and it's performance; in section \ref{sec:bayesian}, we perform mock Bayesian analysis on a case study of Earth and on one hundred random combinations and in section \ref{sec:conclusion} we summarise our findings.

\section{Data Generation}\label{sec:data_gen}

\begin{figure}
\centering
\includegraphics[width=85mm]{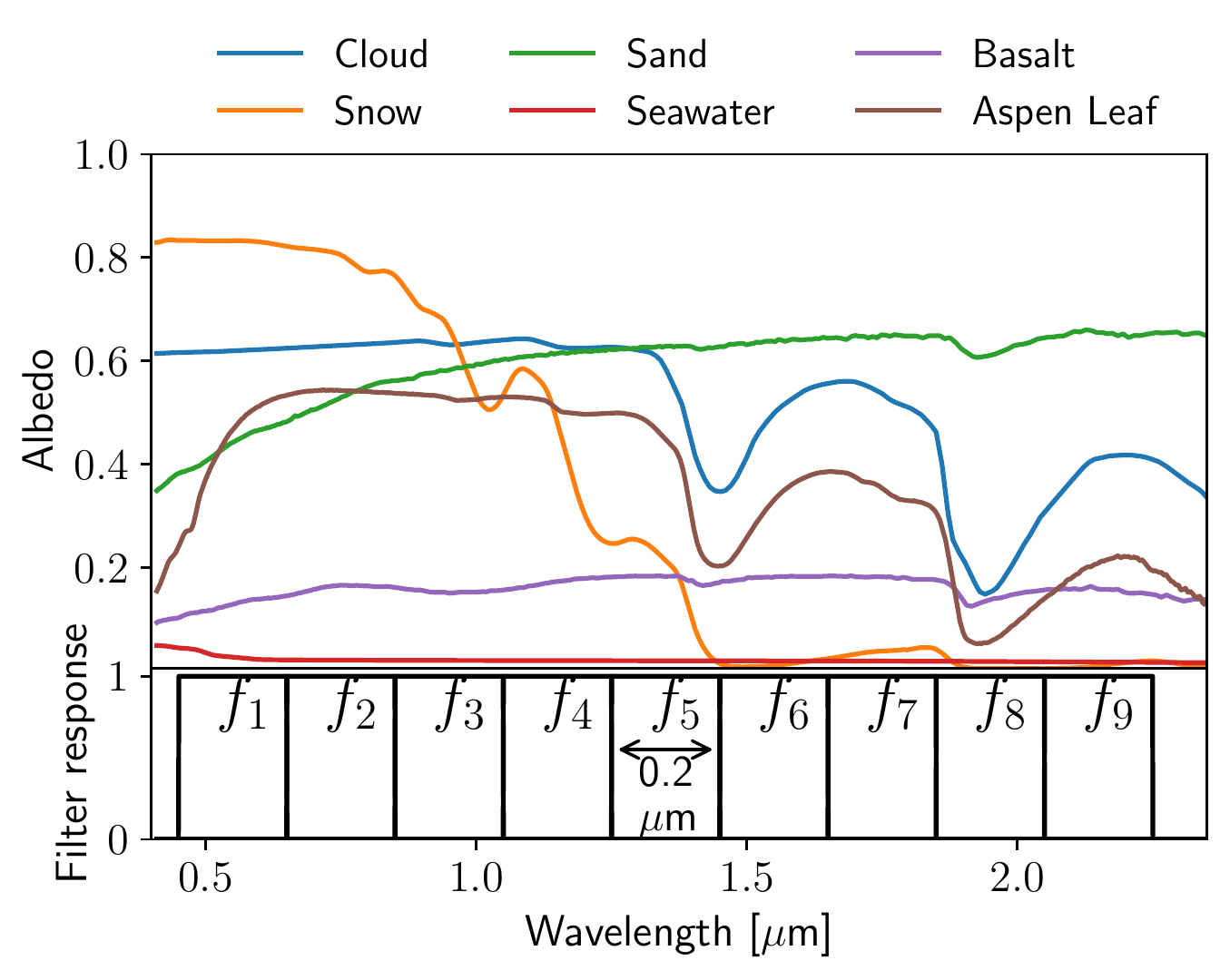}
\caption{\textit{Top}: Albedos of six reflecting components representative of modern Earth, with Aspen leaf as representative for Earth vegetation ( USGS Spectra Library \citep{Clark2003}). \textit{Bottom}: Nine idealised, broadband filters (width = 0.2 $\mu$m) between 0.45 - 2.25 $\mu$m.} \label{fig:components_albedo_filters_response}
\end{figure}

\begin{figure*}
\centering
\includegraphics[width=180mm]{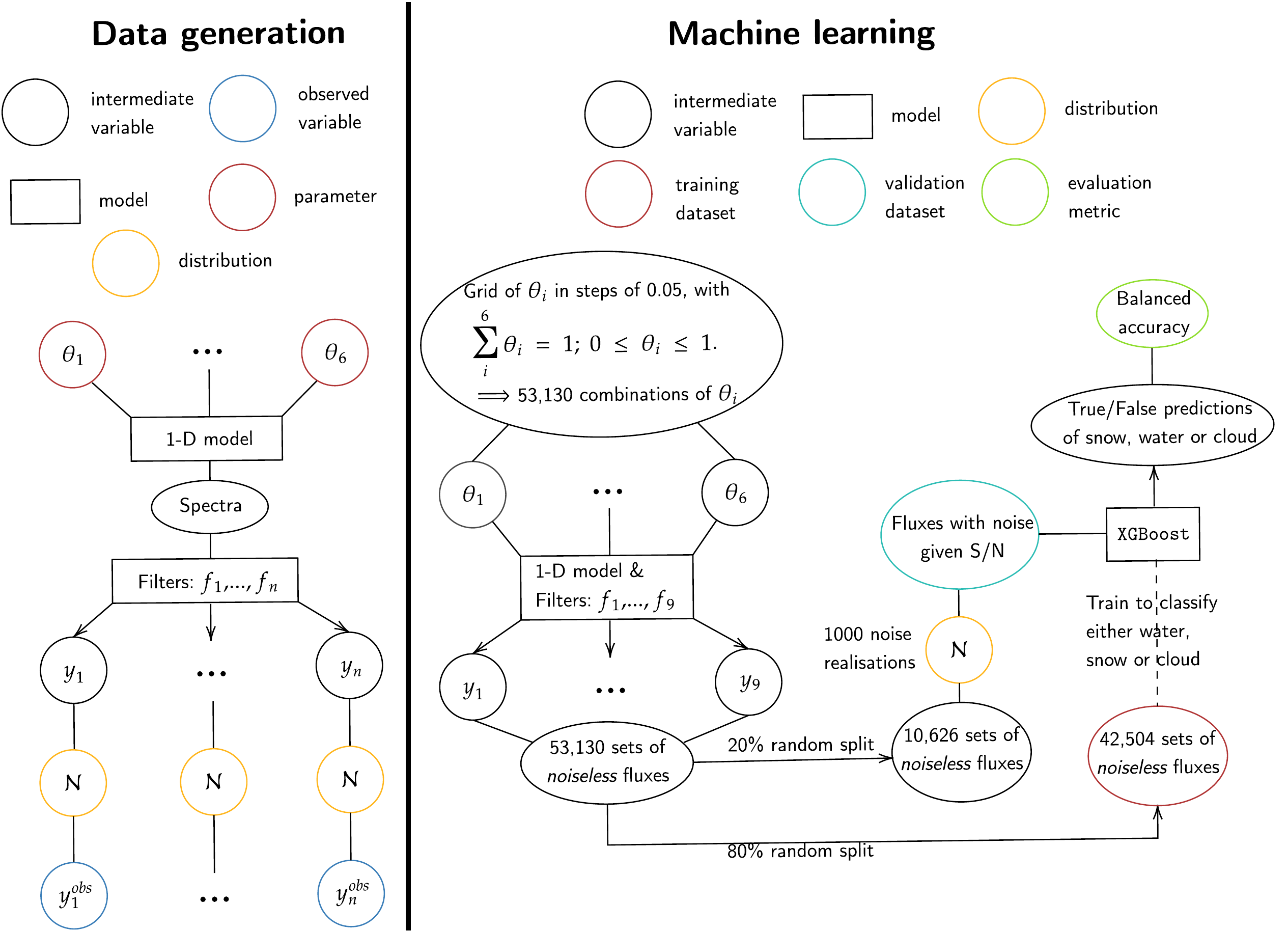}
\caption{\textit{Left}: Graphic representation of the data generation process. $\theta_i$ represents the composition of each reflective component (cloud, snow, etc.), $f_i$ the filters' response, $y_i$ the true integrated flux from each filter and $y^{\mathrm{obs}}_i$ the observed flux. Gaussian noise is added to the true flux ($\mathcal{N}$) producing a signal-to-noise ratio (S/N). \textit{Right}: Graphic of the machine learning pipeline -- data generation, training and validation. $\theta_i$ represents the composition of a particular reflective component (cloud, snow, etc.), $\mathcal{N}$ is the Gaussian model (given some signal-to-noise ratio). To each set of noiseless flux in the 20\% random split noise is added 1000 times (1000 noise realisations). The balanced accuracy is calculated from the prediction of \textit{each} realisation. Hence, at each S/N, we have 1000 balanced accuracy scores. Yields in the shaded 95\% confidence interval are shown in Fig. \ref{fig:balanced_accuracy_feature_importance}.} \label{fig:data_generation_machine_learning}
\end{figure*}

We use a well-known 1D atmospheric model (\texttt{EXO-Prime2}, details in e.g. \citealt{Kaltenegger2021}) to model a slightly colder Earth with a surface temperature of 273K with modern Earth outgassing rates and a Solar spectrum with reduced luminosity of $0.875 L_\odot$ corresponding to a planet further away from the Sun than Earth to generate a cool environment that would allow for snow to cover a large part of the planet. While more complex 3D models are used to model individual planets and explore the effects of surface topography, and rotation rate on specific planets, a computationally cheaper 1D model, which generates a disk integrated flux over an entire planet's surface for broadband filters provides a suitable approximation for the parameter space exploration we undertake here.

We use the six major surface components of modern Earth (water, snow, basalt, vegetation, sand, clouds; following \citealt{Kaltenegger2007}) varying in 5\% steps for surface coverage \citep{PhamKaltenegger2021} to generate the corresponding reflection spectra of 53,130 nominal Earth-size planets. These reflecting components are $\theta_i$, where $i$ ranges from 1-6 representing each component. We create spectra for each nominal planet at a resolution of $0.01~\textrm{cm}^{-1}$ using \texttt{Exo-Prime2}, a 1-D iterative climate-photochemistry code coupled to a line-by-line- radiative transfer code (details in \citealt{Kaltenegger2021, Kaltenegger2007, Kaltenegger2009}). The model has been validated from the visible to infrared through comparison to Earth seen as an exoplanet by missions like the Mars Global Surveyor, EPOXI, multiple Earthshine observations and Shuttle data \citep{Kaltenegger2007,Kaltenegger2009,Rugheimer2013}. It includes the most spectroscopically relevant molecules: C$_2$H$_6$, CH$_4$, CO, CO$_2$, H$_2$CO, H$_2$O, H$_2$O$_2$, H$_2$S, HNO$_3$, HO$_2$, N$_2$O, N$_2$O$_5$, NO$_2$, O$_2$, O$_3$, OCS, OH, SO$_2$, using the HITRAN2016 line lists \citep{Gordon2017} as well as Rayleigh scattering. We divide the planet atmosphere into 52 layers: for each atmospheric layer the code calculates line shapes and widths individually with Doppler- and pressure-broadening with several points per line width.

The other major surface components' albedo are taken from the USGS Spectra Library \citep{Clark2003}, see Fig. \ref{fig:components_albedo_filters_response} with Aspen leaf representative modern Earth vegetation (for the effects of different vegetation and biota and their identification through machine learning on Earth-like planets, see \citealt{PhamKaltenegger2021}) and one opaque water cloud layer at 6 km. The cloud reflectivity \citep[following][]{Madden2020} is based on the MODIS 20 $\mu$m cloud albedo model \citep{King1997,Rossow1999}, which provides an average albedo for clouds of different droplet size and is representative of the 3 main water cloud layers on Earth at 1 km, 6 km and 12 km \citep[see e.g.][]{Kaltenegger2007}. We restrict our spectra wavelength to 0.41 - 2.35 $\mu$m, because these are the available wavelengths of samples from the USGS Spectra Library. Note that we do not adjust the atmosphere profile for different combinations of reflective components, which introduces small changes in the planet's atmosphere and influences their high resolution spectra (see \citealt{Madden2020} for details). Here we are only interested in the broadband fluxes resulting from integrating the spectra over broadband filters, and require tens of thousands of spectra for machine learning and a similar number for model evaluations with Markov-chain Monte Carlo. Hence, keeping a constant atmosphere profile allows us to explore a wide parameter space and evaluate the model efficiently.

We initially created nine, idealised, broadband filters ($f_i$) between 0.45 - 2.25 $\mu$m with a width of 0.2 $\mu$m with an idealised filter response (Fig. \ref{fig:components_albedo_filters_response}) to identify which filters facilitate an effective characterisation of liquid water, snow and clouds on terrestrial exoplanets. We then integrate the model planet spectra with these filters to generate the corresponding true flux, $y_i$, for each filter. We add Gaussian noise to these true fluxes, producing the observed flux $y_i^{\mathrm{obs}}$, given some S/N for our analysis (see Fig. \ref{fig:data_generation_machine_learning} for a summary of the data generation and the machine learning pipeline). For machine learning validation (section \ref{sec:xgboost}), Gaussian noise with S/N in [5, 100] are added. For the Bayesian data generation pipeline (section \ref{sec:bayesian}), Gaussian noise is added at S/N = 10, 50, 100.

\section{Machine Learning with \texttt{XGBoost}}\label{sec:xgboost}

We use \texttt{XGBoost} \citep{Chen2016}, a versatile, gradient boosting algorithm \citep[e.g.][]{Tamayo2016}, to perform binary classification of liquid water, snow and clouds on our model planets. We are classifying if a certain component $\theta_i$ exists (True) or not (False). In our analysis, we train \texttt{XGBoost} to classify only either water, snow or cloud at a time. Through a grid search, we find the default hyperparameters for \texttt{XGBoost} work well for our problem\footnote{We set the \texttt{scale\_pos\_weight} parameter, the ratio of negative to positive samples in the training dataset, but the ratio is fixed for a given dataset.}.

The training and validation dataset are generated like in \citealt{PhamKaltenegger2021}. First, we create a grid of combinations of reflective components $\theta_i$, such that $\sum_{i=1}^{6} \theta_i = 1; 0 \leq \theta_i \leq 1$\footnote{This condition describes the $n-1$ simplex ($n=6$ here). In other words, we create a grid with resolution 0.05 on a 5-simplex.} where each component $\theta_i$ ranges from 0 to 1, in steps of 0.05, subject to the unity sum condition. Therefore, we create a total of 53,130 combinations. Then, we create spectra \footnote{The full spectra and filters dataset is publicly available at \url{https://doi.org/10.5281/zenodo.6234713}.} for each of these combination and nine filter fluxes per spectra. We randomly split this dataset of 53,130 set of nine fluxes into the training and validation datasets at a ratio of 80-20. Thus, the training dataset contains 42,504 sets of fluxes (80\%) and the validation dataset the remaining 10,626 (20\%).

We keep the training dataset noiseless. Gaussian noise is added to the validation dataset for a given S/N. Each validation combination has 1000 noise realisations at a given S/N. This way, we can evaluate our trained model's performance at various noise level. With this methodology, the model is exposed to two different distributions: noiseless for training data and noisy for validation data. Prior work by \cite{Batalha2018, Hayes2020} similarly trained on noiseless data. Our previous work \citep{PhamKaltenegger2021} tested training on noisy and noiseless data and found the latter yielding better results. This approach is versatile because it explores the capability of the algorithms without specifying a noise profile to train on.

To evaluate performance, we use balanced accuracy. The problem resulting from using accuracy instead of balanced accuracy for an imbalanced dataset is discussed in details in \citet{PhamKaltenegger2021}. As a brief summary, a simple metric to evaluating machine learning efficiency is the accuracy (Eq. \ref{eqn:accuracy}). However, this metric only works well if the dataset is balanced -- when sample counts in all classes (True or False in this case) are equally distributed. We label any set of fluxes with (snow, cloud, or water) components at $0\%$ as False, and True otherwise. Our dataset is imbalanced: for example, there are 42,504 combinations with snow (True prediction) yet only 10,626 combinations without snow (False prediction). Using the traditional accuracy usually leads to overly optimistic results \citep{Brodersen2010}. 

For imbalanced datasets, the balanced accuracy should be used instead (Eq. \ref{eqn:balancedaccuracy}), where sensitivity (True Positive Rate) gives the proportion of true positives and specificity (True Negative Rate) gives the proportion of true negatives. Balanced accuracy metric accounts for the true positive and negative rates, instead of averaging over the entire confusion matrix like the accuracy would. Note that another metric that can be used on an imbalanced dataset is a confusion matrix (see \citealt{PhamKaltenegger2021}). However, the balanced accuracy is a scalar value, which can be used when training the model and can easily measure performance quantitatively.

\begin{equation}\label{eqn:accuracy}
    \textrm{accuracy} = \frac{\textrm{correct\ predictions}}{\textrm{total\ predictions}}
\end{equation}

\begin{equation}\label{eqn:balancedaccuracy}
    \textrm{balanced\ accuracy} = \frac{\textrm{sensitivity} + \textrm{specificity}}{2}
\end{equation}

\begin{figure}
\centering
\includegraphics[width=85mm]{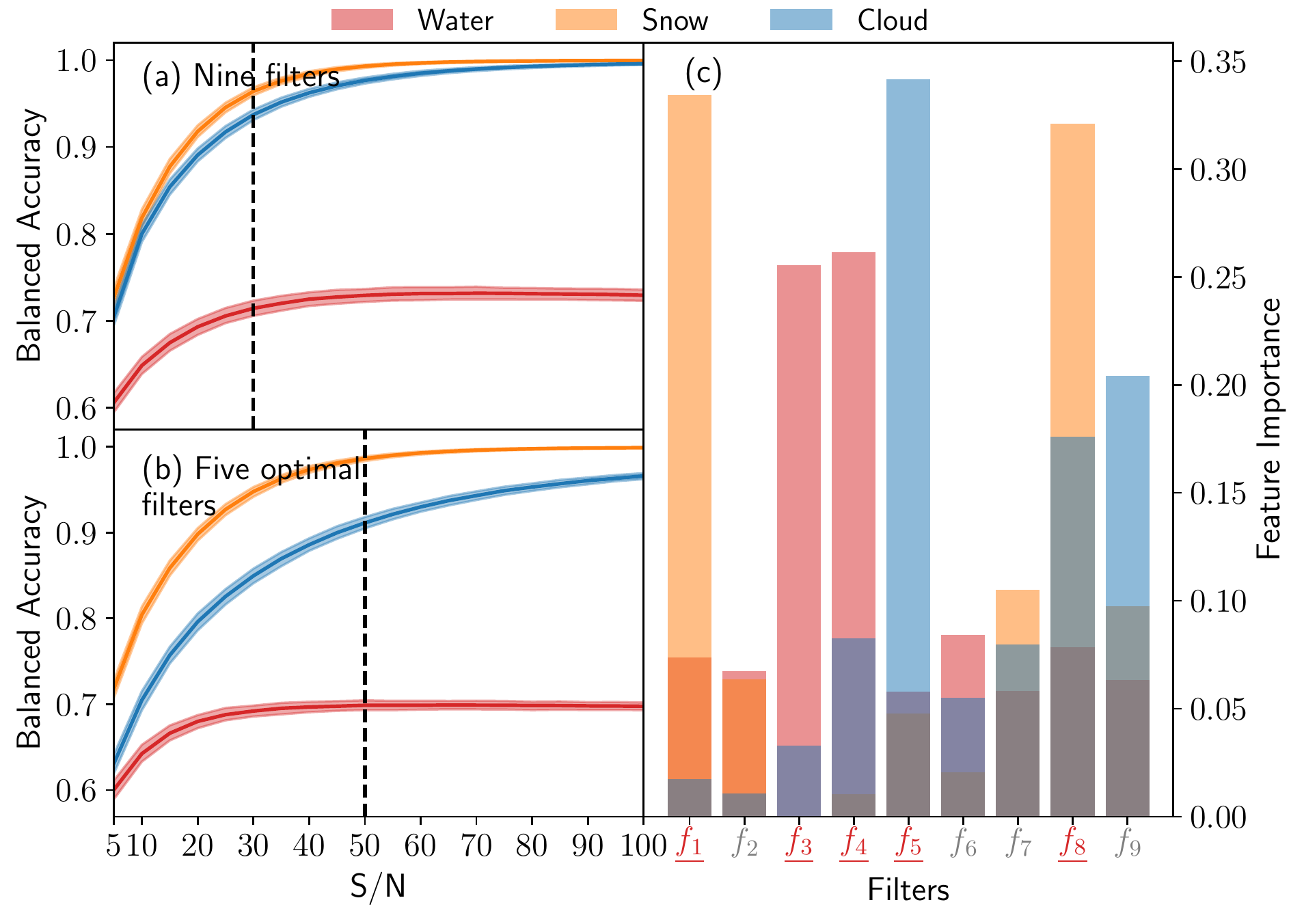}
\caption{\textit{(a)}: Performance (measured with balanced accuracy) of \texttt{XGBoost} at classifying the existence of snow, water or clouds at different signal-to-noise ratio (S/N) with all nine idealised filters. The shaded area represents the 95\% confidence interval from 1000 noise realisations. The algorithm approaches asymptotes at S/N $\approx 30$ (dashed line). \textit{(b)}: Same as (a), but for the five optimal filters. Here, the dashed line is chosen when both water and snow reach their asymptotes and cloud achieve a balanced accuracy $>0.9$.  \textit{(c)}: Feature (filters) importance score ranking. A higher feature score implies greater importance in classification. We select the top five optimal filters (underlined in red).} \label{fig:balanced_accuracy_feature_importance}
\end{figure}

\texttt{XGBoost}'s (trained under nine filters) balanced accuracy over various S/N is shown in panel (a) of Fig. \ref{fig:balanced_accuracy_feature_importance}. Snow and clouds reach asymptotic to approximately unity, water to 0.7 balanced accuracy. Furthermore, the shaded area representing the 95\% confidence interval acquired from 1000 noise realisations is small.

The difference in performance can be explained by the different reflectivity of snow, clouds and liquid water shown in Fig. \ref{fig:components_albedo_filters_response}. The reflectivity of water is low ($<$ 0.1) and has a relatively featureless shape. In contrast, the reflectivity of both snow and clouds show specific features and reflect up to 80\% of the incoming light for snow and up to 60\% for clouds. Liquid water surface features are more difficult to identify than snow or clouds.

Note that Fig. \ref{fig:balanced_accuracy_feature_importance}a shows that for broadband filter photometry a S/N of 20 already provide high accuracy prediction for the existence of water on Earth-like planet models. Beyond S/N = 30, the balanced accuracy becomes asymptotic, meaning additional signal yields a relatively low increase in performance (in a statistical sense).

\texttt{XGBoost} results indicate that machine learning algorithms can be effective tools to classify the existence of snow, clouds and water on Earth-size exoplanets from photometric filter data even at low S/N. 

With the trained model, we identify the set of five optimal filters for identifying water on the surface of an terrestrial exoplanet through photometry between 0.41 - 2.35 $\mu$m using \texttt{XGBoost}'s feature importance score. Here, `features' are the filters for the surfaces we want to identify. Ranking the nine fluxes (as generated by the nine filters) by their importance score allows us to determine the optimal filters (Fig. \ref{fig:balanced_accuracy_feature_importance}c): $f_1$ (0.45 - 0.65), $f_4$ (1.05 - 1.25), $f_5$ (1.25 - 1.45), $f_6$ (1.45 - 1.65), and $f_8$ (1.85 - 2.05). Note that the choice of optimal filters are method (\texttt{XGBoost}) and model dependent (1D atmosphere \texttt{EXO-Prime2}, USGS albedo spectral library). In section \ref{sec:bayesian}, we study if the optimal filters for \texttt{XGBoost} yield effective performance with a Bayesian approach (MCMC).

Comparing the importance scores (Fig. \ref{fig:balanced_accuracy_feature_importance}c) with the wavelength dependent albedo of each component (Fig. \ref{fig:components_albedo_filters_response}) shows peaks of reflectivity of the surfaces in the corresponding filters. The results are based on the built-in XGBoost method, and are consistent with the expectations from surface albedos in Fig \ref{fig:components_albedo_filters_response}. For example, three of the best ranked filters for cloud detection are $f_5$, $f_8$ and $f_9$, which correspond to cloud reflection peaks between 1.6 - 1.9 $\mu$m and 2.1 - 2.3 $\mu$m. Similar correspondence between reflection features and optimal filters exist for snow and water. 

Figure \ref{fig:balanced_accuracy_feature_importance}b shows the balanced accuracy of \texttt{XGBoost} using only the five optimal filters. Water and clouds reach a lower overall balanced accuracy asymptote, snow reaches its earlier asymptote value at higher S/N. These effects are expected because the five filter provide less data compared to the nine filters used in the top panel and our earlier analysis. In Fig. \ref{fig:balanced_accuracy_feature_importance}b, we identify S/N $\approx 50$ (dashed line) as the cutoff for additional signal rewards. However, note that our results show that even a S/N of $\approx 20$ yields a strong performance in accurately predicting the existence of water on the surface of an Earth-like exoplanet.

\section{Bayesian Analysis}\label{sec:bayesian}

We asses the surface composition of our model planets regarding liquid water, snow or clouds with Bayesian inferencing using Markov-chain Monte Carlo (MCMC) for data from the set of five optimal filters identified earlier, That is, inferring $\theta_i$ given $\{y_1^{\mathrm{obs}}, y_3^{\mathrm{obs}}, y_4^{\mathrm{obs}}, y_5^{\mathrm{obs}}, y_7^{\mathrm{obs}}\}$, assuming the model planet spectra and Gaussian noise.

Our posterior distribution function is

\begin{equation}
    \mathcal{P}(\theta|y^{\mathrm{obs}}) \propto \mathcal{U}^*_\theta (0,1) \times \prod_i \frac{1}{\sqrt{2\pi\sigma_i^2}} \exp\left( -\frac{1}{2} \frac{(y_i^\mathrm{obs} - y_i^\theta)^2}{\sigma_i^2} \right) 
\end{equation}

where $\theta = \{\theta_1,...,\theta_6\}$ are the components' composition, $y_i^\mathrm{obs}$ are the (selected optimal) observed fluxes, $y_i^\theta$ are the modelled planetary fluxes in each filter $\theta$, and $\sigma_i$ are the measurement error. The product term is the likelihood function, assuming a Gaussian noise model. Each component $\theta_i \in [0, 1]$, and the sum of all components must be exactly unity. Since we don't know the distribution of exoplanets' surface compositions, we assume that any combination of $\theta_i$ summing to unity is equally likely. Hence, we propose a uniform 5-simplex prior, defined as:

\begin{equation}
    \mathcal{U}^*_\theta (0,1) \propto
    \begin{cases}
        1 & \sum_{i=1}^6 \theta_{i} = 1; 0 \leq \theta_i \leq 1 \\
        0 &\text{otherwise}
    \end{cases}
\end{equation}

We sample this posterior distribution function using \texttt{emcee} \citep{ForemanMackey2013}. 

We first perform a case study of \textit{one} realisation for a model planet with modern Earth's surface composition at S/N = 50 for the five identified filters. This corresponds to a sample case of a planet-by-planet analysis that would be done for a specific exoplanet (here we choose the Earth model case). Modern Earth surface seen from space consists on average of about 50\% clouds. The 50\% surface is composed of about 70\% seawater, 5.4\% basalt, 4.5\% snow, 2.1\% sand, and 18\% vegetation \citep{Kaltenegger2007}. These factors represent our $\theta_i$'s for this case, which sum to unity.

\begin{figure}
\centering
\includegraphics[width=50mm]{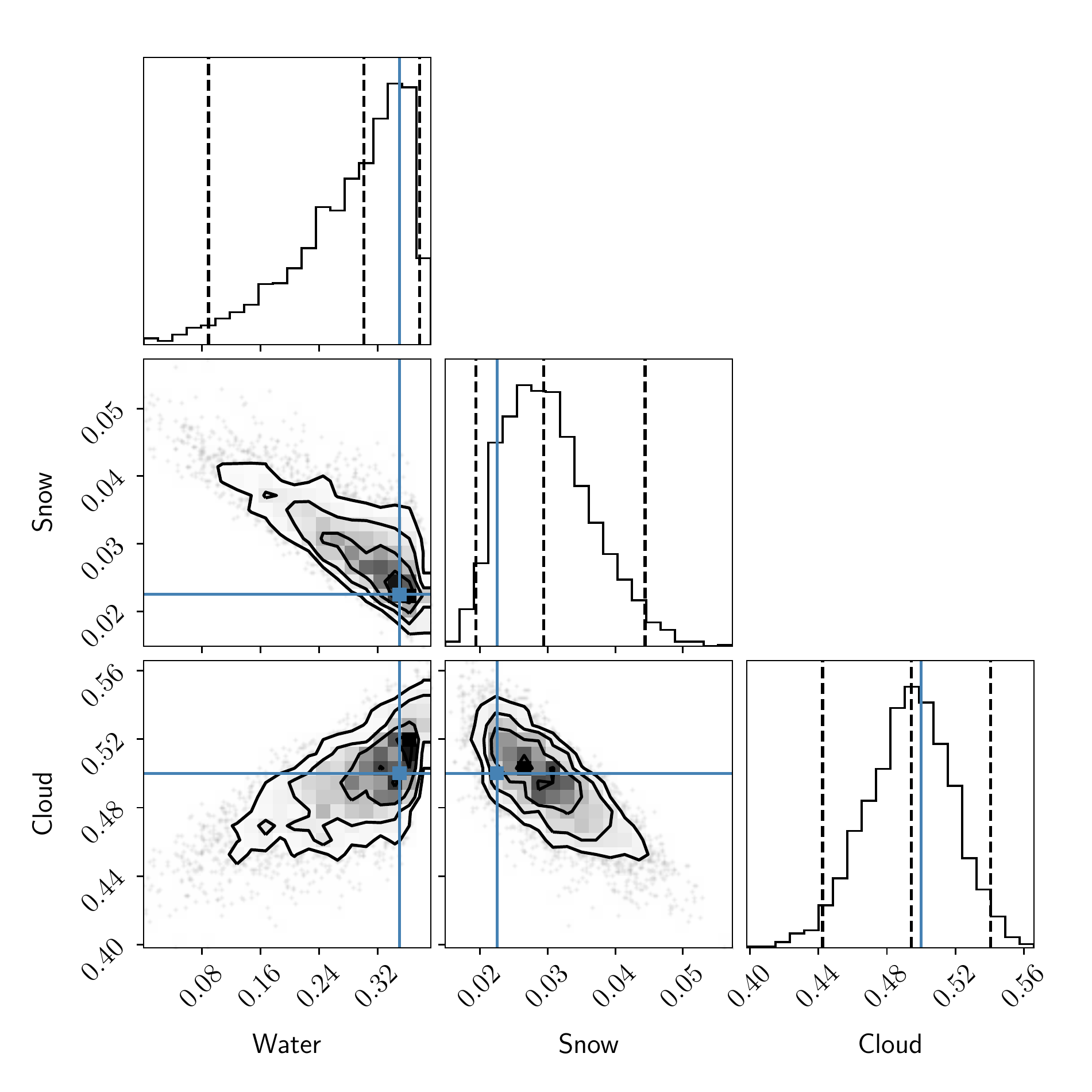}
\caption{Corner plot showing the 1-D and 2-D marginalised distributions of water, snow and clouds for one random realisation with modern Earth surface composition at S/N = 50. The dashed lines represent the median and the 95\% credible interval, the solid blue lines mark the true value.} \label{fig:MCMC_corner_Earth_estimates}
\end{figure}

Figure \ref{fig:MCMC_corner_Earth_estimates} shows the 1-D and 2-D marginalised distributions of water, snow and clouds for this one particular realisation at S/N = 50: It shows the median and 95\% credible interval estimates from MCMC and the true composition. The MCMC's 95\% credible interval covers the true value, and the median estimate is close as well. The results demonstrate that it is feasible to infer the surface composition regarding water as liquid, snow and clouds on a terrestrial exoplanet based on the proposed five optimal filters, within a 95\% credible interval for this particular combination. 

\begin{figure}
\centering
\includegraphics[width=85mm]{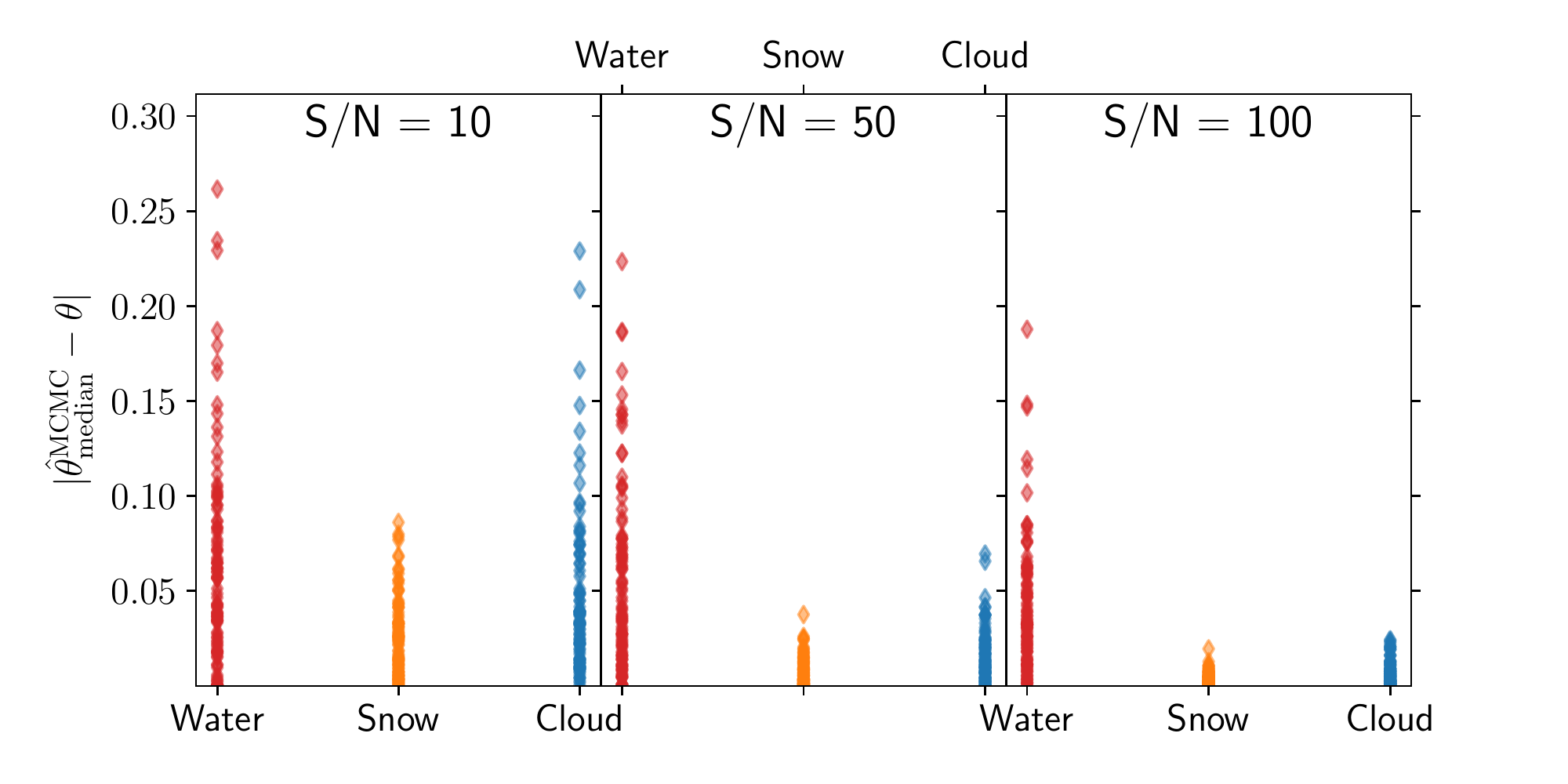}\caption{Residual between MCMC median estimates and the true values at S/N = 10, 50, 100, for one hundred random compositions. $\hat{\theta}^{\mathrm{MCMC}}_{\mathrm{median}}$ is the median estimate from MCMC and $\theta$ is the true value. The greatest residual for all three S/N is $\max | \hat{\theta}^{\mathrm{MCMC}}_{\mathrm{median}} - \theta| = 0.26$. }
\label{fig:MCMC_random_residual}
\end{figure}

Figure \ref{fig:MCMC_random_residual} shows the results when we extend the analysis to one hundred random combinations, each with added noise at S/N of 10, 50 and 100. It shows the residual between MCMC median estimates, $\hat{\theta}^{\mathrm{MCMC}}_{\mathrm{median}}$, and the true values, $\theta$. Note that the residual for snow and cloud reduces significantly as S/N increases. This corresponds to the balanced accuracy performance from \texttt{XGBoost} with increasing S/N (Fig. \ref{fig:balanced_accuracy_feature_importance}) due to the reflection features of the three surfaces (see Fig. \ref{fig:components_albedo_filters_response}). Snow and clouds reflect more incoming star light and show features over the wavelength range covered. In contrast the albedo of liquid seawater is comparably smooth, making liquid water more difficult to identify than snow or clouds. Finally, most residuals are accurate to within 5\% for snow and clouds at S/N $\gtrsim 50$ -- these result highlighting the potential of using photometric filters to assess the existence of water on a terrestrial exoplanet.

\section{Conclusion}\label{sec:conclusion}

We explored the feasibility of detecting water on the surface of terrestrial exoplanets in its various forms (snow, clouds and liquid water) through broadband photometry using machine learning and Markov-chain Monte Carlo (MCMC).

First, we trained a well known, versatile machine learning algorithm, \texttt{XGBoost}, to perform binary classification on the existence of snow, clouds and water on an exoplanet's surface using broadband photometric flux. The performance show promise to use machine learning on photometric data to identify water on the surface of terrestrial exoplanets (Fig. \ref{fig:balanced_accuracy_feature_importance}a,b): the algorithm achieves a high ($> 90\%$) balanced accuracy for snow and clouds for S/N $\gtrsim 20$) and up to 70\% balanced accuracy for liquid water.

Second, we identified five optimal filters to identify snow, clouds and liquid water on a terrestrial planet's surface between 0.45 - 2.25 $\mu$m 
based on \texttt{XGBoost}'s feature importance ranking. These optimal filters (Fig. \ref{fig:balanced_accuracy_feature_importance}c), could be implemented in telescopes designs that search for water on exoplanets.

Third, we tested these optimal filters by performing Bayesian inferences using MCMC on i) an Earth case study, and ii) one hundred random realisations of our planetary models. The results show promise, with most predictions at S/N $\gtrsim 50$ within 5\% of the true value for finding snow and clouds. Detecting liquid water is more challenging, but most predictions are within 20\% of the true values.

In a typical workflow, we expect to use machine learning for fast prioritisation and use MCMC for further constraints. The advantage of machine learning is it's fast run time (in our runs, about 1 minute for 53,000 sets of fluxes), comparing to MCMC sampling's (about 15 minutes per set of fluxes). We find that machine learning can be an additional fast and accurate retrieval pathway. Note that MCMC would only require $\sim 2500$ hours for 10,000 samples.

The results showed that broadband filter photometry combined with machine learning provides a promising tool for initial characterisation of water in different forms on the surface of terrestrial exoplanets and their prioritisation for time-intense follow-up observations for a smaller observing time cost. 

Broadband filter photometry could allow large missions like HabEx and LUVOIR to initially characterise targets for follow-up and smaller telescopes like the Nancy Grace Roman telescope to help with initial characterisation of nearby rocky worlds.

\section*{Acknowledgements}

We thank the anonymous reviewer for their insightful suggestions that have improved the clarity and strengthened our paper. We would like to thank David Ruppert, Joshua S. Speagle, Samantha Berek and Zifan Lin for helpful discussions and comments.  


\section*{Data Availability}

The data underlying this article are available in \url{https://doi.org/10.5281/zenodo.6234713}.

\bibliographystyle{mnras}
\bibliography{citations}{}

\bsp	
\label{lastpage}
\end{document}